\documentclass[aps,prl,twocolumn,showpacs]{revtex4-1}
\usepackage{graphicx}
\usepackage{bm}
\usepackage{wasysym}
\usepackage{setspace}

\begin{document}

%\preprint{\sf Version 1 (\today)}
\title{Deconfinement of electric charges in hydrogen-bonded ferroelectrics}
\author{Bo-Jie Huang}
\author{Chyh-Hong Chern}
\email{chchern@ntu.edu.tw}
\affiliation{Department of Physics, National Taiwan University, Taipei 10617, Taiwan}
\date{\today}
\begin{abstract}
In addition to the gauge charges, a new charge degree of freedom is found in the deconfined phase in the lattice Ising gauge theory.  While applying to the hydrogen-bonded ferroelectrics, the new charge is essentially the electric charge, leading to the divergent dielectric susceptibility.  The new degree of freedom paves an experimentally accessible way to identify the deconfined phase in the lattice Ising gauge theory.
 \end{abstract}

\maketitle

%\setstretch{2.0}

Deconfinement is a basic notion in the lattice gauge theory~\cite{Kogut, Savit}.  When the system is in the decondined phase, it takes only finite amount of energy to separate the \emph{gauge charges} to infinite distance.  There is no local order parameter to distinguish the confinement and the deconfinement.  It is the non-local Wilson loop which behaves differently in two phases.  Namely, it obeys an area law in the confined phase and a perimeter law in the deconfined phase.  The property of the non-locality makes the deconfined phase intractable to be identified in real systems.

Recently, one of us (CHC) proposed that an Ising gauge theory can be realized in the hydrogen-bonded ferroelectrics obeying the ice rule, for example KH$_2$PO$_4$ (KDP) and squaric acid (H$_2$SQ)~\cite{Chern}.  Ferroelectric (FE) materials have been applied in many devices.  In particular, the ferroelectric organic hydrogen-bonded systems obtain more and more attentions because of the advantages of flexibility, light-weightiness, and non-toxicity.

Besides the ice~\cite{Pauling}, the ice rule was widely observed in many frustrated magnetic systems, particularly in the three dimensional pyrochlore systems~\cite{Bramwell, Nisoli}.  The rule constrains the spin dynamics.  In each tetrahedron of the pyrochlore lattice, spins form a "two-in-two-out" configuration.  Namely, two of the four spins point inward to the center of the tetrahedron, and the other two point outward simultaneously.  Following this rule, there could be a large number of ground state configurations.  The macroscopic ground state degeneracy is essential for those systems to host spin disorder ground state, which intrigues people in searching for new exotic states.  

The similar configuration is also observed in KH$_2$PO$_4$ (KDP) and squaric acid (H$_2$SQ)~\cite{Slater, quantum1, quantum2, quantum3, SQacid}.  The KDP and H$_2$SQ molecules are planer molecules with four corners, where two of them are occupied by hydrogen ions.  As the molecules form a corner-shared two-dimensional network, the hydrogen ions can tunnel between the nearest-neighboured molecules. Then, the ice rule is applied so that there are only two hydrogen ions in each molecule. In addition to the ice rule, the two hydrogen ions always occupy at the nearest-neighboured corners.  In other words, two hydrogen ions occupying at the diagonal corners is never allowed.  Interpreting the tunnelling of the hydrogen dynamics as a gauge degree of freedom, a lattice gauge theory can be constructed to implement the ice rule in those materials.  Consequently, different from a conventional FE transition, the one in those hydrogen-bonded systems is close to a confinement-deconfinement phase transition~\cite{Chern}.

The confinement-deconfinement phase transition is a hallmark transition in the lattice gauge theory.  Local order parameters do not exist to distinguish phases since any quantity without gauge invariance has zero ground state expectation value.  The gauge invariant quantity is, however, often non-local in nature.  Since there is no order parameter associated with measurable quantities, direct measurement and verification is very difficult.  Nevertheless, it was suggested that the dielectric susceptibility, defined by
\begin{eqnarray}
\chi = \frac{1}{N}\sum_{i,j}\int_0^{\beta}d\tau<P_x(i, \tau)P_x(j,0)>, \label{dieX}
\end{eqnarray}
could be an indirect tool to identify the deconfined phase in those ferroelectric systems, where $P_x$ is the $x$-component polarization vector, $N$ is the number of molecules, and $\beta=1/(k_BT)$~\cite{Chern}.   At zero temperature, the dielectric susceptibility diverges in the deconfined phase and obeys the Currie-Weiss-like law in $K$ (defined later) perfectly in the confined phase.  

The dielectric susceptibility measures the fluctuation of the electric charge degree of freedom.  However, the electric charges in those systems have nothing to do with the gauge charges.  Although the deconfined phase of the gauge charges exists in the lattice Ising gauge theory, it is unknown whether other new deconfined phase exists or not, in additional to the gauge charges.  Inspired by the divergent dielectric susceptibility found in Ref.~\cite{Chern}, we shall show that a new deconfined phase, corresponding to the electric charges in the ferroelectrics, exists in the lattice Ising gauge theory.  Not only the non-local Wilson loop but also the gauge charge in the lattice Ising gauge theory do not have experimental relevancy.  Our study provides an experimentally accessible way to identify the phase of the deconfinement.

\begin{figure}[htb]
\includegraphics[width=0.45\textwidth]{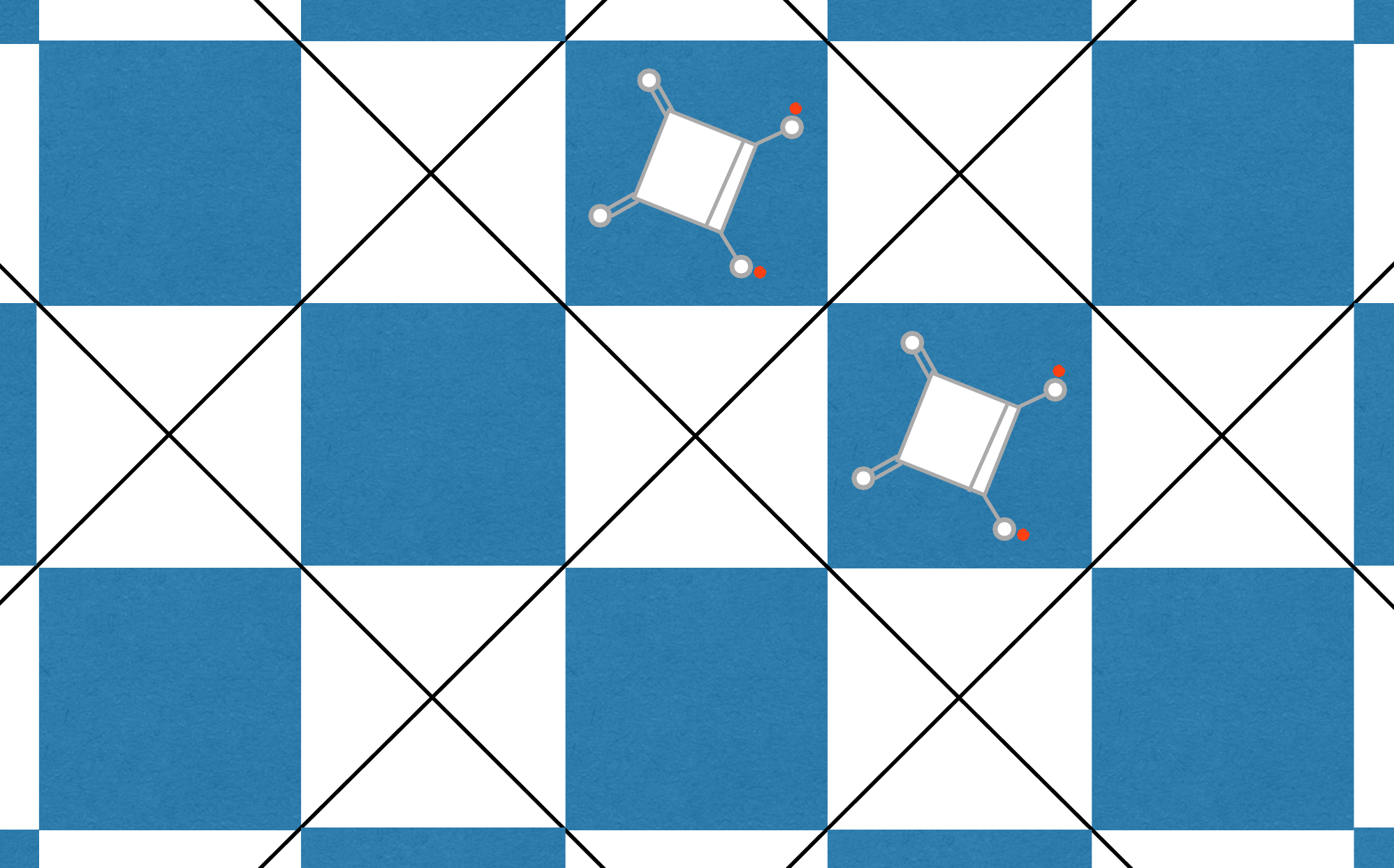}
\caption{(Color online) The two dimensional corner-shared network of molecules.  Blue squares denote the molecules.  Black lines represent the lattice.  Two squaric acid molecules are drawn as an example.  The red dots denote the hydrogen ions.}\label{Fig:lattice}
\end{figure}

Let us begin with briefly reviewing the modelling of the hydrogen-bonded ferroelectrics, especially the property of the dielectric susceptibility.  The two-dimensional corner-shared array of KDP or squaric acid molecules is shown schematically in Fig.~(\ref{Fig:lattice}).  At each connection corner, two states can be defined to represent the positions of the hydrogen ions.  Using the Pauli matrix $\sigma^z$, the Hamiltonian can be written as
\begin{eqnarray}
\!\!\!\!H = &-&J_0\sum_{\Box}\sigma^z_1\sigma^z_2\sigma^z_3\sigma^z_4+J_1\sum_{\Box}(\sigma^z_1\sigma^z_3+\sigma^z_2\sigma^z_4)\nonumber\\ &-&J_2\!\sum_{<AB>}\!\vec{P}_A\cdot\vec{P}_B-K\sum_{i}\sigma^x_i, \label{h}
\end{eqnarray}
where $\Box$ runs all molecules and $\sigma^x$ is the tunnelling matrix of the hydrogen ions.  $\vec{P}_{(A, B)}$ are the polarization vectors for the bipartite $A$ and $B$ molecules, defined by $P_{(A, B)x}=(\pm)\frac{1}{4}(\sigma^z_1+\sigma^z_2-\sigma^z_3-\sigma^z_4)$ and $P_{(A, B)y}=(\pm)\frac{1}{4}(\sigma^z_2+\sigma^z_3-\sigma^z_1-\sigma^z_4)$, where $(+)$ for the molecule $A$ and $(-)$ for the molecule $B$ respectively.  The Hamiltonian in Eq.~(\ref{h}) was already considered in Ref.~\cite{Chern}.  It can be understood in the following way.  The $J_0$ and $J_1$ terms impose the ice rule and that no hydrogen ions can occupy diagonal to each other.  In most cases, $J_1$ is much smaller than $J_0$.  The $J_2$ term introduces the ferroelectric interaction.  The $K$ term allows the hydrogen to switch positions quantum mechanically between nearest-neighboured molecules.  

When $J_2$ is finite and $K$ is small, the ground state is a ferroelectric state.  A second-order phase transition takes place at some $K_c(J_2)$, a function of $J_2$, to take the system to the dielectric phase.  It is a conventional second-order ferroelectric transition, since the dielectric susceptibility $\chi$ well satisfies the Curie-Weiss-like behaviour $\chi=C/(K-K_c)$ for $K > K_c$.  When $J_2$ is zero, the system is in the deconfined phase for small $K$.  Another second-order phase transition takes the system from the deconfined phase to the confined phase at some $K_{\text{CDT}}$, which is called the confinement-deconfinement phase transition.  Interestingly, the ferroelectric transition evolves smoothly to the confinement-deconfinement phase transition with $K_{\text{CDT}} = K_c(J_2=0)$.  In other words, at zero temperature, the ferroelectric phase is bordered by the deconfined phase, and the dielectric phase is bordered by the confined phase.

The deconfined phase is a disorder phase, since the polarization vector is not a gauge invariant quantity and should have the zero ground-state expectation value.  Not only the polarization but also the spatial correlation function $<\vec{P}(r)\vec{P}(0)>\ \ \propto \ \ \frac{e^{-r/\xi_{\text{FT}}}}{r}$ is zero with $\xi_{\text{FT}} \sim \sqrt{J_2}$ at small $J_2$, where $\xi_{\text{FT}}$ is the ferroelectric correlation length.  Furthermore, the dielectric susceptibility diverges in the deconfined phase at zero temperature, since the integrand in Eq.~(\ref{dieX}) has a power-law relation in $\tau$ when $i=j$.  However, the divergent behaviour of the dielectric susceptibility has nothing to do with the deconfinement of the gauge charges which are defined on the vertices in the lattice.  Nevertheless, it sheds light on the possibility of the new deconfinement in the lattice Ising gauge theory.  In the following, we shall demonstrate that the new deconfinement is the one of the electric charges in the hydrogen-bonded ferroelectrics.  

To begin with, let us focus on the $J_0$ and the $K$ term
\begin{eqnarray}
\!\!\!\!H_0 = &-&J_0\sum_{\Box}\sigma^z_1\sigma^z_2\sigma^z_3\sigma^z_4-K\sum_{i}\sigma^x_i, \label{gauge}
\end{eqnarray}
which is nothing but the two-dimensional quantum Hamiltonian of the three-dimensional Ising gauge theory.
A second order critical point at $K_c=0.325J_0$ separates the confined phase and the deconfined phase~\cite{Blote}.  When $K=0$, Eq.~(\ref{gauge}) reduces to a classical model with infinite degenerate ground states, since for each $\Box$ there are 8 configurations in the minimization of the $J_0$ term.  It is also the two-dimensional Ising gauge theory which is dual to the one-dimensional Ising model.  The $K$ term is a singular perturbation, which introduces an additional dimension and is essentially a single flip which takes the system away from the ground state.  The excitation of the single flip is costly in energy, which takes $2J_0$ in the zero$^{\text{th}}$ order approximation.  On the other hand, if we perform a collective four-flips, where the four $\sigma^z$ share the same vertex, the collective four-flips brings the system from one ground state to another one with a different configuration.  It lifts the ground state degeneracy and results in a large number of low-energy excitations.  The effective theory can be obtained by performing a fourth order perturbation theory, and we found
\begin{eqnarray}
H'_0 = -J_0\sum_{\Box}\sigma^z_1\sigma^z_2\sigma^z_3\sigma^z_4-J_e\sum_{+}\sigma^x_\alpha\sigma^x_\beta\sigma^x_\gamma\sigma^x_\delta, \label{eq4}
\end{eqnarray}
where $J_e=\frac{5K^4}{16J_0^3}$, and $+$ runs on all vertices~\cite{Terhal}.  The Eq.~(\ref{eq4}) is nothing but the renowned Kitaev's toric code model~\cite{Kitaev1, Kitaev2}.

\begin{figure}[htb]
\includegraphics[width=0.45\textwidth]{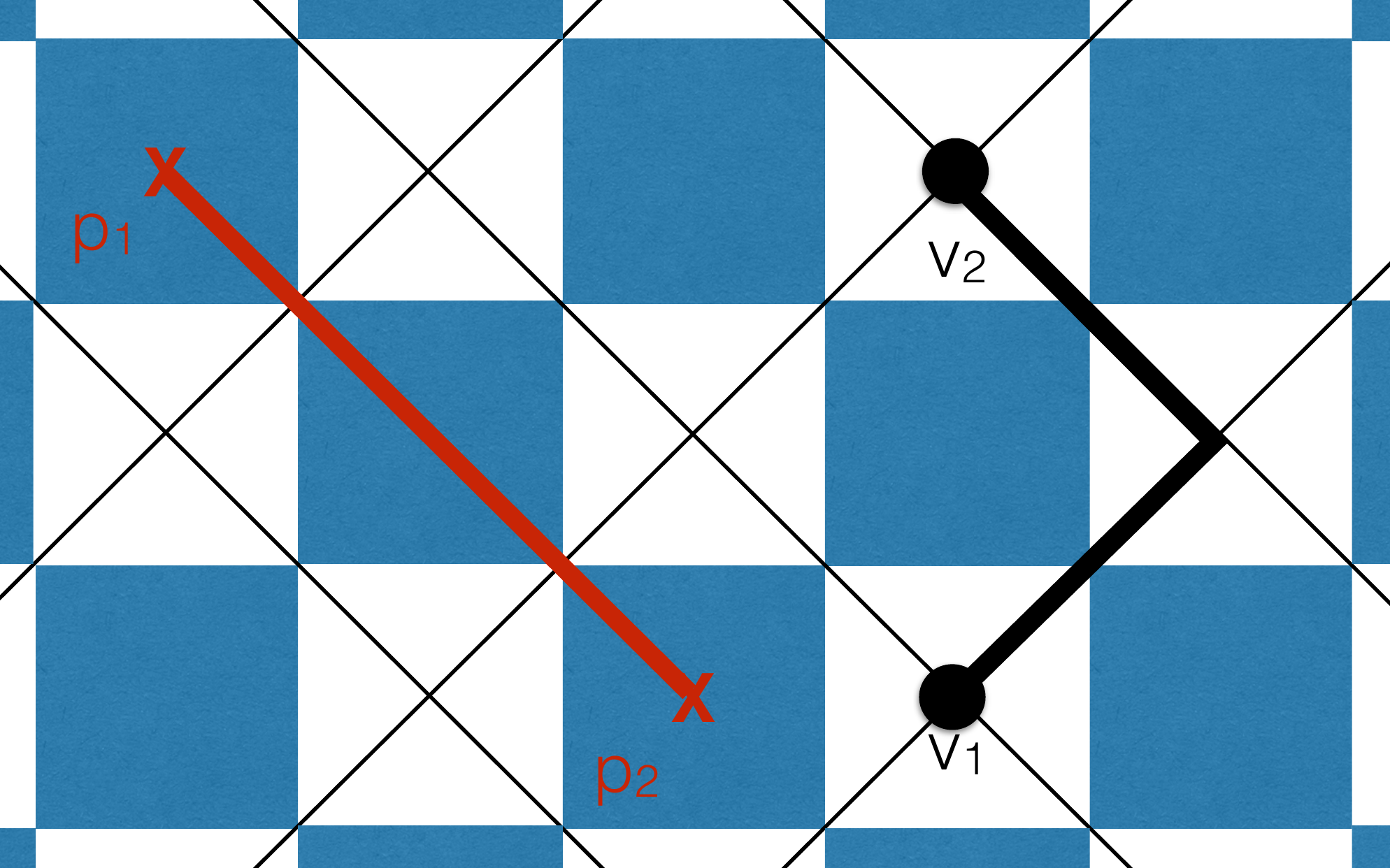}
\caption{(Color online) The operator $e$ creates two gauge charges of the $\eta$ variables at vertices $v_1$ and $v_2$ with a string of length $l=2$.  The operator $m$ breaks the electric neutrality of the molecules at the end points $p_1$ and $p_2$ and creates two charged molecules with a string of length $l^*=2$.}\label{Fig:excitation}
\end{figure}

The Kitaev model is an exact soluble model with topological ground state degeneracy.  The exact excitations of the Kitaev model can be defined by the string operator $e$
\begin{eqnarray}
W^{(e)}_l = \prod _{j \in l} \sigma^z_j,
\end{eqnarray}
and the string operator $m$
\begin{eqnarray}
W^{(m)}_{l^*} = \prod_{j\in l^*} \sigma^x_{j},
\end{eqnarray}
and their fusion $e\times m$.  In Fig.~(\ref{Fig:excitation}), two examples of the excitations are shown.  One can define a $\eta$ variable, valued at $+1$ or $-1$, on the vertex and define $\sigma^z_j =\eta_i\eta_j$ on the bond between the nearest-neighboured vertices.   The $e$ excitation is simply $W^{(e)}_l = \eta_i\eta_f$, where $i$ and $f$ denote the starting and the end points of the string respectively.  Thus, the $\eta$ variable is nothing but the gauge charge.  The energy of the $e$ excitation is $4J_e$, which is actually very small in our case.  Since $K_c=0.325J_0$, the maximum $J_e$ is equal to $0.001133J_0$. On the other hand, the $m$ excitation is a string of switching hydrogen positions.  When applying $W^{(m)}_{l^*}$ to the ground state, the molecules at the ends of the string beak the electric neutrality.  Therefore, the $m$ excitation introduces a pair of electric charge excitations and is relevant to the dielectric susceptibility.  The energy of the $m$ excitation is $4J_0$.  It can be easily seen that the energy of both $m$ and $e$ excitations do not depend on the lengths of strings, that represents the signature of the deconfinement. 

In the real ferroelectrics, $J_1$ is not zero and restricts every molecule to satisfy the ice rule.  Numerical study indeed demonstrates the divergent dielectric susceptibility at finite $J_1$ and zero $J_2$. Next, we should discuss the gauge charge excitations $e$ and the electric charge excitation $m$ under this condition.  Defining $H'_1=H'_0+J_1\sum_{\Box}(\sigma^z_1\sigma^z_3+\sigma^z_2\sigma^z_4)$ and $B_{p}=\sum_{\Box_p}\sigma^z_1\sigma^z_2\sigma^z_3\sigma^z_4$, the energy of the charge excitation $m$ can be computed as
\begin{eqnarray}
&&\Delta E_m=<\Phi|W^{(m)}_{l^*} H'_1W^{(m)}_{l^*} |\Phi>-<\Phi|H'_1|\Phi> \label{energyM}\\
&&=<\Phi|W^{(m)}_{l^*} [H'_1, W^{(m)}_{l^*}]|\Phi> \nonumber \\
&&=2J\!<\!\Phi|(B_{p_i}\!\!+\!B_{p_f})|\Phi\!>\!-2J_1\!\!<\!\Phi|(\sigma^z_{1i}\sigma^z_{3i}\!+\!\sigma^z_{1f}\sigma^z_{3f})|\Phi\!>, \nonumber
\end{eqnarray}
where $|\Phi>$ is the ground state of the $H'_1$, and the subscripts $i$ and $f$ represent the end molecules of the string operator.  Eq.~(\ref{energyM}) suggests that electric charge excitations do not depend on the separation distance and are thus deconfined.  In the real system, $J_1$ term should be much bigger than $J_e$ term, and they do not commute with each other.  It is possible to compute $\Delta E_m$ in the perturbation theory in $J_e/J_1$.  In the first order perturbation, we obtain
\begin{eqnarray}
\Delta E_m = 4J_0+4J_1-\frac{J_e^2}{8J_1}.
\end{eqnarray}
Nevertheless, in the presence of $J_1$ term, the charge excitation $m$ is no longer an exact excited state.  However, using the  properties, 
\begin{eqnarray}
W^{(m)}_{l^*}B_p &=& -B_p W^{(m)}_{l^*} \nonumber \\
B_p|\Phi>&=&|\Phi> \ \ \ \forall \ p, \label{bp}
\end{eqnarray}
it can be shown that the excited state $W^{(m)}_{l^*} |\Phi>$ has zero overlap with the ground state, namely $<\Phi|W^{(m)}_{l^*}|\Phi>=0$.  These results suggest that the charge fluctuations are robust and deconfined, leading to a divergent behaviour of the dielectric susceptibility.  

Similarly, the energy of the $e$ excitation can be computed as $\Delta E_e = 2J_e<\Phi|(A_{s_i}+A_{s_f})|\Phi>=\frac{J_e^2}{2J_1}$ in the first order perturbation.  $\Delta E_e$ does not depend on the length of the string, either.  Likewise, the $e$ excitation has zero overlap with the ground state due to the hidden gauge symmetry.  The analytical approximation given above provides additional supports that the deconfined phase extends to the finite $J_1$ region, consistent with the previous study.

Finally, we consider the full account of the Hamiltonian, Eq.~(\ref{eq4}) plus the $J_1$ and $J_2$ terms in Eq.~(\ref{h}).  The energy of the charge excitation $m$ is $\Delta E_m=4J_0+4J_1-\frac{J_e^2}{8J_1}+2J_2(l^*+1)$.  Numerical calculation confirms that the ground state is ferroelectric at any finite $J_2$.  Similar to a string excitation in a magnetically-ordered system, the energy of the charge excitation $m$, a string of switching hydrogen ions, is proportional to the length of the string $l^*$.  This property is known as the confinement.  

We note that the second equation in Eq.~(\ref{bp}) is not true if we consider the original Hamiltonian with the $K$-term in Eq.~(\ref{gauge}).  Since the Hamiltonian in Eq.~(\ref{gauge}) breaks the time-reversal symmetry, at finite $K$, the ground-state expectation value of the $W^{(m)}_{l^*}$ operator is not zero, which in principle weakens the stability of the charge excitation $m$.  Therefore, the current results focus on the situation when the $K$-term is small and can be analyzed effectively by the $J_e$-term in Eq.~(\ref{eq4}).

The above analysis demonstrates that the low-energy effective theory of three-dimensional Ising gauge theory in the cubic lattice is the Kitaev's toric-code model.  Due to the exactness of the Kitaev's model, the spectrum confirms the existence of the deconfinement of two charge excitations.  In addition to the well-known gauge charge, we found a new deconfined excitation corresponding the electric charge which breaks the electric neutrality of the molecule.  The gauge charges on the $\eta$ variable are defined on the vertices, which are the empty space between molecules.  Therefore, the gauge charges loses any experimental relevancy.  The electric charges, however, is properly defined by the gauge field.  Namely, the operator $\sigma^z_1\sigma^z_2\sigma^z_3\sigma^z_4$ of a single molecule measures the total electric charge of that molecule, which is much easier to be measured in experiments.

The dielectric susceptibility is the most relevant quantity associated with the electric charges.  It simply reflects the ability to induce the dipole moment in responding to the external electric field.  The deconfinement of the electric charge implies a divergent dielectric susceptibility.  The analysis from above extends the deconfinement further to the finite $J_1$, which is the experimentally relevant region.  Now the question is how to achieve to the region, namely zero $J_2$?  Due to the shape of the molecules, the effective ferroelectric interaction of the $J_2$ term is caused by the \emph{steric hindrance}~\cite{Ishizuka}.  It is an effect of geometrical frustration, which helps all molecules align.  As every molecule carries a finite dipole moment, the alignment results in a ferroelectric phase.  Because of the soft flexibility of these materials, one can stretch them mechanically to make more space for molecules.  The enlargement of the space reduces $J_2$ and eventually reaches zero $J_2$.  The stretching process reduces not only $J_2$ but also $K$.  Since the distance between molecules increases, the matrix element for the hydrogen switching is further reduced, which hence increases $T_c$~\cite{Tatsuzaki}.  As experiments are always conducted at finite temperature, the infinite dielectric susceptibility will never be observed.  At finite temperature, dielectric susceptibility is much enhanced but finite in the deconfined phase comparing to in the ferroelectric phase, since Eq.~(\ref{dieX}) becomes a definite integral.

Finally, we conclude that by mapping to the Kitaev's model, we found a new deconfined charge degree of freedom.  Unlike the non-local Wilson loop and the gauge charge, the new charge can be expressed by the gauge field and finds its experimental relevancy as the electric charge in the hydrogen-bonded ferroelectrics.  The deconfinement of the new charge leads to the divergent dielectric susceptibility, which was numerically confirmed in Ref.~\cite{Chern}.  One of the most beautiful property of the three-dimensional Ising gauge theory is the duality of the three-dimensional Ising model.  Although the exact solution of the 3D Ising model is still difficult, our approach might stimulate some progress along that direction, relying on the exactness of the Kitaev's model.  As the Kitaev's model is proposed for the quantum computation, its relevancy to the hydrogen-bonded ferroelectrics might find itself possible to be realised in those materials.  On the other hand, although the lattice gauge theories have been studied for many decades, most of them remain inapplicable to real systems.  There might be some hidden charge deconfinement existing in the abelian or the non-abelian gauge theory, which are relevant to experiments~\cite{Nisoli}.  Exploring new structures can be promising directions of future research.

CHC acknowledges the fruitful discussion with Naoto Nagaosa and Ching-Chou Tsai.  This work is supported by Ministry of Science and Technology of Taiwan under the grant: MOST 103-2112-M-002-014-MY3 and by National Taiwan University under the grant: 103R7831 and 104R7831.

%\bibliography{FMIR4}
%Merlin.mbs v4.21 2009-07-09.

\end{document}